\documentstyle{aipproc}
\newcommand{\bq}{\begin{equation}}
\newcommand{\eq}{\end{equation}}

\newcommand{\gsim}{\raisebox{-0.07cm}{$\, \stackrel{>}{{\scriptstyle
\sim}}\, $}}

\newcommand\GeV{\,\mbox{GeV}}

\begin{document}

\begin{flushleft}
DESY 97--105\\
{\tt hep-ph/9706362}
\end{flushleft}

\vspace*{-8mm}
\title{On the Leptoquark Interpretation of the High $Q^2$ Events at
HERA}

\author{Johannes Bl\"umlein}
\address{DESY-Zeuthen, Platanenallee 6, D-15735~Zeuthen,
Germany }

\maketitle

\begin{abstract}
We investigate to which extent an interpretation of the recently
observed excess of events in the  high $Q^2$ range at HERA in terms of
single leptoquark production is compatible with bounds from other
experiments.
\end{abstract}

\vspace*{-5mm}
\section*{Introduction}
\noindent
Recently both the H1~\cite{J:H1} and ZEUS~\cite{J:ZEUS} experiments
reported an excess of deep inelastic neutral current events in the range
$Q^2 \gsim 15000 \GeV^2$. There is also a slight indication of an excess
for charged current scattering in the H1 data, cf.~\cite{J:H1}. Various
phenomenological and theoretical
analyses~\cite{J:REF,J:PPSS,J:JB,J:QED2,J:QCD1,J:BK}
were performed shortly after
this finding to seek an interpretation, though the experimental
signals have first to stabilize adding in more data in the current runs.
One possible interpretation for these events might be single leptoquark
production in the $e^+ q$ or $e^+\overline{q}$ channels, respectively.
In this note we summarize the status of the phenomenological studies
which were performed during the last months.
\footnote{
For surveys on the earlier literature on leptoquarks
see e.g. refs.~\cite{J:JB,J:BB}.}
Other interpretations,
such as a signal for substructure~\cite{J:SUB}, effects due to R--parity
violating supersymmetry~\cite{J:SUSY},
and others~\cite{J:OTH}, as well as implications for experiments at
$e^+e^-$ colliders~\cite{J:EPM}, were also discussed.

\vspace*{-5mm}
\section*{$p\overline{p}$ Scattering}
\noindent
Leptoquarks may be searched for at hadron colliders both studying
single and pair production processes. In the case of the
single production
processes~\cite{J:PSING} the reaction cross sections are
$\propto  \lambda^2$ and amount to
$\sigma_{\rm sing} \sim 0.4~fb ... 1.3~fb$
at Tevatron energies
for  $(\lambda/e)\sqrt{Br}  \approx 0.075$, eq.~(\ref{J:E:HB}),
only, \cite{J:JB}, and are too small to be detected currently.

On the other hand, given the small fermionic couplings, the pair
production processes~\cite{J:SCA,J:VEC1,J:VEC2,J:BBK} depend on
the leptoquark--gluon couplings only. In the case of scalar leptoquarks
the production cross section is completely predicted, whereas it
depends on anomalous couplings, such as  $\kappa_G$ and $\lambda_G$, in
the case of vector leptoquarks. As was shown in ref.~\cite{J:BBK},
however, there exists a global minimum ${\sf min}_{\kappa_G, \lambda_G}
\left [\sigma(\Phi_V \overline{\Phi}_V)\right]  > 0$
allowing for a model-independent analysis.

The production cross sections for scalar  leptoquarks in the
partonic subsystem read~\cite{J:SCA,J:BBK}
\begin{eqnarray}
\sigma_{\Phi_S \overline{\Phi}_S}^{q \overline{q}} &=&
\frac{2 \pi \alpha_s^2}{27 \hat{s}} \beta^3\\
\sigma_{\Phi_S \overline{\Phi}_S}^{gg} &=&
 \frac{\pi \alpha_s^2}{
96 \hat{s}} \left [ \beta (41 - 31 \beta^2) - (17 - 18 \beta^2 + \beta^4)
\ln \left|\frac{1+\beta}{1-\beta}\right| \right],
\end{eqnarray}
with $\beta = \sqrt{1 - 4 M_{\Phi}^2/\hat{s}}$. The $O(\alpha_s)$
correction  to the production cross section was calculated
in ref.~\cite{J:PPSS} and amounts to a $K$-factor of 1.12 only
for the choice
$\mu = M_{\Phi}$ of  the factorization scale.
The cross sections
 in the case of vector leptoquark pair production
are more complicated, cf.~\cite{J:BBK}, due to the
presence of the anomalous couplings $\kappa_G$ and $\lambda_G$ and have
the general structure
\begin{eqnarray}
\sigma_{\Phi_V \overline{\Phi}_V}^{q \overline{q}} &=&
\frac{4 \pi \alpha_s^2}{9 M_V} \sum_{i=0}^5 \chi_i^q(\kappa_G,\lambda_G)
\widetilde{G}_i(\hat{s}, \beta)\\
\sigma_{\Phi_V \overline{\Phi}_V}^{gg} &=&
\frac{\pi \alpha_s^2}{96 M_V} \sum_{i=0}^{14}
\chi_i^g(\kappa_G,\lambda_G)
\widetilde{F}_i(\hat{s}, \beta)~.
\end{eqnarray}
The functions $\chi^{q,g},\widetilde{G}_i$ and $\widetilde{F}_i$ are
given in ref.~\cite{J:BBK}.
For $\kappa_G = \lambda_G = 0$ one obtains~\cite{J:VEC1,J:BBK}
\begin{eqnarray}
\sigma_{\Phi_V \overline{\Phi}_V}^{q \overline{q}} =
\frac{ \pi \alpha_s^2}{54 M_V} \left[ \frac{\hat{s}}{M_V^2}
+ 23 - 3 \beta^2 \right],
\sigma_{\Phi_V \overline{\Phi}_V}^{gg} =
 \frac{\pi \alpha_s^2}{96 M_V} \left\{\beta
A(\beta) - B(\beta)
\ln \left|\frac{1+\beta}{1-\beta}\right| \right \},
\nonumber
\end{eqnarray}
\begin{eqnarray}
A(\beta) = \frac{523}{4} - 90 \beta^2 + \frac{93}{4} \beta^4,~~~~~
B(\beta) = \frac{3}{4} \left[
65 - 83 \beta^2 + 19 \beta^4 - \beta^6 \right]~.
\end{eqnarray}
Choosing the factorization and renormalization scales by
$\mu = M_{\Phi}$
the pair production cross sections   for scalar and vector
leptoquarks (minimizing for $\kappa_G$ and $\lambda_G$)
at Born level
and using the parametrization~\cite{J:CTEQ3}
for the parton densities are~\cite{J:JB}~:
\begin{eqnarray}
\label{J:E:CS}
\sigma_S(M_{\Phi} = 200 \GeV) = 0.16~pb~~~~~~~\sigma_V(M_{\Phi} =
200 \GeV) = 0.29~pb.
\end{eqnarray}

\vspace*{-5mm}
\section*{Excluded Mass Ranges}
\noindent
The most stringent limits on leptoquark masses, which are independent of
the fermionic couplings, were derived by the Tevatron experiments,
see~\cite{J:TEV}, searching for leptoquark pair
production~\cite{J:SCA,J:VEC1,J:VEC2,J:BBK,J:PPSS}.
The following  mass ranges
are excluded for scalar
leptoquarks associated to the first fermion
generation~:
\begin{eqnarray}
\label{J:E:TL}
 M  <~~~~~~~210~\GeV & & {\rm~CDF}~~~Br(eq) = 1\nonumber\\
 M  < 176~(225) \GeV & & {\rm~D0}~~~~~~Br(eq) = 0.5~(1)
\end{eqnarray}
at 95\%~CL.
The mass bounds for vector leptoquarks are correspondingly higher because
of the larger production cross section, eq.~(\ref{J:E:CS}), but have not
yet been presented by the Tevatron experiments for the general
case~\cite{J:BBK}.

\vspace*{-5mm}
\section*{The HERA Events}
\noindent
If the observed high-$Q^2$ excess is interpreted in terms of single
leptoquark production~\cite{J:EP1,J:BRW,J:EP3}
 constraints on the fermionic coupling $\lambda$
of the leptoquarks $\Phi$, which may be either scalars or vectors,
may be derived. Due to the location of the excess found by H1 in
the range $M = \sqrt{x S}  \sim 200~\GeV$ we  assume this scale
in the estimates given below.
In the narrow width approximation the production cross
section reads
\begin{equation}
\sigma = \frac{\pi^2}{2} \alpha \left ( \frac{\lambda}{e} \right )^2
q(x, \langle Q^2 \rangle) \left \{\begin{array}{c} 2~:~V\\ 1~:~S
\end{array} \right. \times Br(\Phi \rightarrow e q)~.
\end{equation}
For the observed excess in the $e^+ + jet$ channel at H1
\begin{equation}
\label{J:E:HB}
\frac{\lambda_S}{e} \sqrt{Br}
\sim 0.075~~(0.15)~~~~u~~(d),~~~\Phi = S
\end{equation}
is derived~\cite{J:JB} using the parametrization~\cite{J:CTEQ3} of
the quark densities, while $\lambda_V = \lambda_S/\sqrt{2}$
and $\lambda_{\rm ZEUS} = 0.55 \lambda_{\rm H1}$. These couplings are
well compatible with the limits derived from low energy
data~\cite{J:LE}. For the production cross section of scalar leptoquarks
the QCD corrections were calculated in \cite{J:QCD1}. They amount to
$+ 23\%$. Both for the measurement of the total cross section and
differential distributions,
such as the mass distribution $M = \sqrt{xS}$
or the $y$ distribution, a precise treatment of the QED radiative
corrections is of imortance, because these corrections can be very large
depending on the way in which the kinematic variables are
measured, see ref.~\cite{J:QED1} for a general discussion. The universal
QED corrections due to initial and final state radiation can be
calculated in the leading log approximation, accounting also for higher
orders using the code {\tt HECTOR}~\cite{J:QED1}. A similar study
was  performed in \cite{J:QED2}.

An information on the spin of the produced state
can be derived from the $y$ distribution of the events. The statistics
is yet to low to allow for such an analysis. The average value $\langle
y \rangle_{\rm H1} = 0.59 \pm 0.02$ is still compatible with both the
expectation for a scalar $\langle y \rangle_S = 0.65$ or a vector
$\langle y \rangle_V = 0.55$, cf.~\cite{J:JB}.
One also may consider the scattering process $e q \rightarrow \Phi g$
both for scalar and vector
leptoquarks~$\Phi$~\cite{J:EP1,J:BK}\footnote{A few events with
$e + 2 jet$ final states
have been already observed in the high $Q^2$ range
at HERA~\cite{J:TC}.}. For this process the angular distributions
are different in the case of scalars  and vectors,
cf. ref.~\cite{J:BK} for details.

A severe constraint on the leptoquark states which may be produced
in $e^+ q$ scattering is imposed by the $SU(2)_L \times U(1)_Y$ quantum
numbers\footnote{For a classification in the case of family-diagonal,
baryon- and lepton number conserving, non-derivative couplings, see
\cite{J:BRW}.}.  If besides the $e^+ q$ final states the indication
of also $\nu q$ final states becomes manifest, scalar leptoquarks
are {\bf not} allowed since low energy constraints demand either
$\lambda_L \ll \lambda_R$ or $\lambda_R \ll \lambda_L$. For the vector
states $U^0_{3\mu}$ or $U_{1\mu}$, which may be produced
in the $e^+ d$ channel, on the other hand, the branching ratios
are $Br(e^+ d) = Br(\nu u) = 1/2$.

The bounds in eq.~(\ref{J:E:TL}) exclude both a scalar and a vector
leptoquark with a mass of $M \sim 200 \GeV$ and $Br(eq) =1$ at 95\% CL.
These bounds are widely model independent.
If $Br(eq) = 0.5$, the limits derived at  the Tevatron and the
interpretation of the excess of events in the high $Q^2$ range at HERA
in terms of single leptoquark production are still compatible.
It will be interesting to see, whether also an excess of $\nu jet$
final states is observed by the ZEUS experiment in the high $Q^2$ range.
If the excess of events persists in the data taken throughout this
year, a first determination of
the $\nu jet/e^+ jet$ ratio will be possible.


\vspace*{-5mm}
\begin{references}
%
\bibitem{J:H1}
C. Adloff et al., H1 collaboration, {\it Z. Phys.} {\bf C74} 191
(1997).
%
\bibitem{J:ZEUS}
J. Breitweg et al., ZEUS collaboration, {\it Z. Phys.} {\bf C74} 207
(1997).
%
\bibitem{J:REF}
S.L. Adler, IASSNS-HEP-96/104; {\tt hep-ph/9702387};
D. Choudhury and S. Raychaudhuri, {\tt hep-ph/9702392};
T. Kuo and T. Lee, {\tt hep-ph/9703255};
G. Altarelli et al., {\tt hep-ph/9703276};
H. Dreiner  and P. Morawitz,  {\tt hep-ph/9703279};
M.A. Doncheski and S. Godfrey,  {\tt hep-ph/9703285};
J. Kalinowski et al.,  {\tt hep-ph/9703288} and {\it Z. Phys.} {\bf C}
in print;
K. Babu et al.,  {\tt hep-ph/9703299};
V. Barger et al., {\tt hep-ph/9703311};
M. Suzuki, {\tt hep-ph/9703316};
M. Drees, {\tt hep-ph/9703332};
J. Hewett and T. Rizzo,  {\tt hep-ph/9703337};
G.K. Leontaris and J.D. Vergados,  {\tt hep-ph/9703338};
M.C. Garcia and S.D. Novaes,  {\tt hep-ph/9703346};
D. Choudhury and S. Raychaudhuri, {\tt hep-ph/9703369};
C. Papadopoulos,
                {\tt hep-ph/9703372};
N. Di Bartolomeo and M. Fabbrichesi, {\tt hep-ph/9703375};
A. Nelson, {\tt hep-ph/9703379};
J. Kalinowski et al.,  {\tt hep-ph/9703436};
B. Arbuzov, {\tt hep-ph/9703460};
C. Friberg et  al.,  {\tt hep-ph/9704214};
T. Kon and T. Kobayashi, {\tt hep-ph/9704221};
S. Jadach et al., {\tt hep-ph/9704241};
A. White, {\tt hep-ph/9704248};
R. Barbieri et al., {\tt hep-ph/9704275};
I. Montvay, {\tt hep-ph/9704280};
W. Buchm\"uller and D. Wyler, {\tt hep-ph/9704317};
K. Akama et al., {\tt hep-ph/9704327};
S. Kuhlmann et al., {\tt hep-ph/9704338};
S. King and G. Leontaris, {\tt hep-ph/9704336};
G. Giudice  and R. Rattazzi, {\tt hep-ph/9704339};
A. Belyaev and A. Gladyshev, {\tt hep-ph/9704343};
J. Elwood et al., {\tt hep-ph/9704363};
S. Godfrey, {\tt hep-ph/9704380};
M. Chizov, {\tt hep-ph/9704409};
B. Dutta et al., {\tt hep-ph/9704428};
M. Heyssler and W. Stirling, {\tt hep-ph/9705229};
N. Deshpande et al., {\tt hep-ph/9705236};
G. Altarelli et al., {\tt hep-ph/9705287};
S. Barshay and G. Keyerhoff, {\tt hep-ph/9705303};
D. Roy et al., {\tt hep-ph/9705370};
K. Babu et al., {\tt hep-ph/0705399};
{\tt hep-ph/0705414};
J. Ellis et al., {\tt hep-ph/9705416};
A. Deandrea, {\tt hep-ph/9705435}
%
\bibitem{J:PPSS}
M. Kr\"amer et al., {\tt hep-ph/9704322}.
%
\bibitem{J:JB}
J. Bl\"umlein, {\tt hep-ph/9703287} and {\it Z. Phys.} {\bf C}
in print.
%
\bibitem{J:QED2}
S. Jadach et al., {\tt hep-ph/9705395}.
%
\bibitem{J:QCD1}
Z. Kunszt and W. Stirling,  {\tt hep-ph/9703427};
T. Plehn et al.,  {\tt hep-ph/9703433}.
%
\bibitem{J:BK}
J. Bl\"umlein and A. Kryukov, DESY 97--067.
%
\bibitem{J:BB}
J. Bl\"umlein and E. Boos, Nucl. Phys. {\bf B} (Proc. Suppl.)
{\bf 37B} 181 (1994).
%
\bibitem{J:SUB}
D. Zeppenfeld, these proceedings.
%
\bibitem{J:SUSY}
S. Lola, these proceedings.
%
\bibitem{J:OTH}
S. Kuhlmann, these proceedings; A. White, these proceedings.
%
\bibitem{J:EPM}
J. Kalinowski, these proceedings, {\tt hep-ph/9706203}.
%
\bibitem{J:PSING}
A. Dobado et al., {\it Phys. Lett.}  {\bf B207} 97 (1988);
J. Hewett and S. Pakvasa, {\it Phys. Rev.} {\bf D37} 3165 (1988);
J. Cieza--Montalvo and O. Eboli, {\it  Phys. Rev.} {\bf D50} 331
(1994); A. Djouadi et al., SLAC--PUB--95--6772;
J. Ohnemus et al., {\it Phys. Lett.} {\bf B334} 203 (1994); E. Reya,
private communication.
%
%
\bibitem{J:SCA}
J.A. Grifols and A. M\'endez, {\it Phys. Rev.} {\bf D26} 324 (1982);
P.R. Harrison and C.H. Llewellyn Smith, {\it Nucl. Phys.} {\bf B213} 223
(1983); Erratum: {\bf B223} 524 (1983); I. Antoniadis et al.,
{\it Z. Phys.} {\bf C23} 119 (1984); E. Eichten  et al., {\it Rev. Mod.
Phys.} {\bf 56} 579 (1984); G. Altarelli and R. R\"uckl,
{\it Phys. Lett.} {\bf B144} 126 (1984); S. Dawson et al.,
{\it Phys. Rev.} {\bf D31} 1581.
%
\bibitem{J:VEC1}
P. Arnold and C. Wendt, {\it Phys. Rev.} {\bf D33} 1873 (1986);
G.V. Borisov et al., {\it Z. Phys.}  {\bf C36} 217 (1987).
%
\bibitem{J:VEC2}
J.L. Hewett  et al., in: Proc. of the Workshop {\sf Physics at Current
Accelerators and  Supercolliders}, eds. J.L.~Hewett
et al.,  (ANL, Argonne, 1993), 539; T.G. Rizzo, {\tt hep-ph/9609267}.
%
\bibitem{J:BBK}
J. Bl\"umlein, E. Boos, and A. Kryukov,
{\tt hep-ph/9610408}, {\it Z. Phys.}  {\bf C} in print.
%
\bibitem{J:CTEQ3}
H.L.~Lai et al., {\it Phys. Rev.} {\bf D51} 4763 (1995).
%
\bibitem{J:TEV}
H.S. Kambara, CDF collaboration, Hadron Collider Physics XI, June 5-11
1997;\\
G. Landsberg, D0 collaboration, Seminar, Fermilab, June 6 1997;
G. Wang, these proceedings.
%
\bibitem{J:EP1}
J. Wudka, Phys. Lett. {\bf B167} (1986) 337.
%
\bibitem{J:BRW}
W. Buchm\"uller, R. R\"uckl, and D. Wyler, {\it Phys. Lett.} {\bf B191}
(1987) 442.
%
\bibitem{J:EP3}
A. Dobado, M.J. Herrero, and C. Mu\~{n}oz, Phys. Lett. {\bf B191} (1987)
449.
%
\bibitem{J:LE}
W. Buchm\"uller and D. Wyler, {\it Phys. Lett.} {\bf B177} 377 (1986);
M. Leurer, {\it Phys. Rev. Lett.} {\bf 71} 1324 (1993); {\it Phys. Rev.}
{\bf D49} 333 (1994); {\it Phys. Rev.} {\bf D50} 536 (1994);
S. Davidson et al., {\it Z. Phys.}  {\bf C61} 613 (1994).
%
\bibitem{J:QED1}
A. Arbuzov et al., {\tt HECTOR 1.00},
{\tt hep-ph/9510410}, {\it Comp. Phys. Commun.} {\bf 94} 128
(1996).
%
\bibitem{J:TC}
T. Carli, private communication.
\end{references}
\end{document}